\documentclass[pdflatex,sn-mathphys]{sn-jnl}

\usepackage{lmodern}
\usepackage[superscript]{cite}
\usepackage{amsfonts,amssymb,mathtools}

\jyear{2022}%

\theoremstyle{thmstyleone}%
\theoremstyle{thmstyletwo}%

\theoremstyle{thmstylethree}%

\begin{document}

\title[Disruptive Changes in Field Equation Modeling]{%
\begin{center}
Disruptive Changes in Field Equation Modeling \\ 
\large  A Simple Interface for Wafer Scale Engines
\end{center}
}



\author[1,2]{\fnm{Mino} \sur{Woo}}

\author[1]{\fnm{Terry} \sur{Jordan}}

\author[3]{\fnm{Robert} \sur{Schreiber}}

\author[3]{\fnm{Ilya} \sur{Sharapov}}

\author[3]{\fnm{Shaheer} \sur{Muhammad}}

\author[3]{\fnm{Abhishek} \sur{Koneru}}

\author*[3]{\fnm{Michael} \sur{James}}\email{michael@cerebras.net}

\author*[1]{\fnm{Dirk} \sur{Van Essendelft}}\email{dirk.vanessendelft@netl.doe.gov}

\affil[1]{ \orgname{National Energy Technology Laboratory}, \orgaddress{\city{Morgantown}, \postcode{26505}, \state{WV}, \country{United States}}}

\affil[2]{\orgname{Oak Ridge Institute for Science and Education}, \orgaddress{\city{Oak Ridge}, \postcode{37830}, \state{TN}, \country{United States}}}

\affil[3]{\orgname{Cerebras Systems Inc}, \orgaddress{\city{Sunnyvale}, \postcode{94085}, \state{CA}, \country{United States}}}




\abstract{We present a high-level and accessible Application Programming Interface (API) for the solution of field equations on the Cerebras Systems Wafer-Scale Engine (WSE) with over two orders of magnitude performance gain relative to traditional distributed computing approaches. The domain-specific API is called the WSE Field-equation API (WFA). The WFA outperforms OpenFOAM\textsuperscript{\textregistered} on NETL's Joule 2.0 supercomputer by over two orders of magnitude in time to solution. While this performance is consistent with hand-optimized assembly codes, the WFA provides an easy-to-use, high-level Python interface that allows users to form and solve field equations effortlessly.  We report here the WFA programming methodology and achieved performance on the latest generation of WSE, the CS-2.}





\maketitle

\section{Introduction}\label{sec1}

According to the principle of locality, an object is directly influenced by its immediate surroundings. Outside of limited quantum phenomena,\cite{bib4,bib5} the principal of locality holds true for fundamental forces \cite{braibant2011particles} and is the foundation of a mathematical construct called a field. A field represents any physical quantity that has a value at every point in spacetime.\cite{feynman1989} Examples include gravitational and electromagnetic fields.  The principle of locality and the field concept underpin sets of partial differential equations (known as field equations), which describe most phenomena observed around us.

\begin{figure}[h]%
\centering
\includegraphics[width=1.0\textwidth]{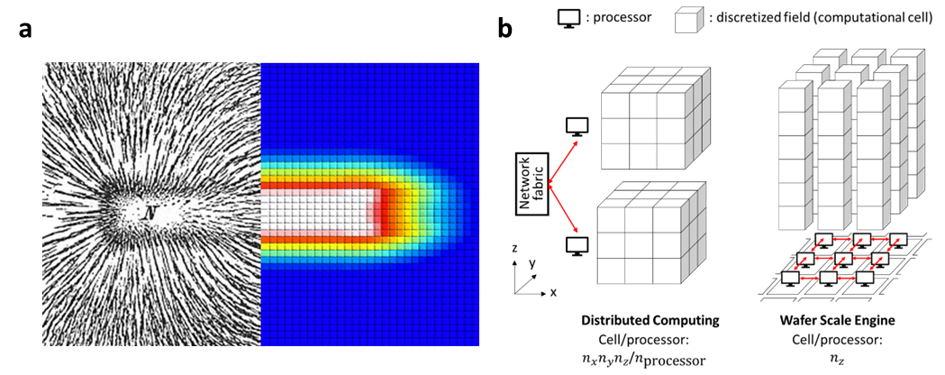}
\caption{(a) A magnetic field\cite{bib2} (left) and its approximation through discretization in a field equation computer model (right). (b) The parallel computation strategy of the current generation of distributed computing and WSE.}\label{fig1}
\end{figure}

To solve these equations on computers, practitioners divide space into discrete control volumes (often called cells or voxels) where each field is approximated by a single average value (Fig \ref{fig1}a). The terms \emph{grid} and \emph{mesh} refer to the entire set of control volumes. Differences between neighboring cells provide approximations of the field's gradients.  Field equations specify sparse relations among field values and field gradients that hold over all spacetime. 

Computations that accurately model complex physical phenomena can be very large, requiring a prohibitively long time to solve, unless high-performance computing (HPC) systems comprising thousands of processing nodes are employed.  To distribute the computation between multiple nodes, practitioners apply domain decomposition methods\cite{hoffmann1993computational} with individual nodes solving the problem over distinct physical regions (Fig. \ref{fig1}b, left). 

Applying computational resources in parallel increases the performance, measured as the amount of aggregate computational work divided by the time taken to solve the problem.  There are two common strategies to scale workloads on a distributed system.  First, strong scaling, which keeps the problem size in the numerator fixed while applying additional processor nodes, decreases the time to solution.  Second, in contrast with weak scaling, the problem size increases proportionally with the node count, while the resulting time to solution in the denominator stays roughly constant.\footnote{Total time to solution will only remain constant if cell resolution is constant as problem size grows.  For CFD simulations, increasing cell resolution may exceed the Courant--Friedrichs--Lewy limit, which could require more iterations per unit runtime.  In addition, for explicit formulations, the stability may be affected which may require reduced time step size.} 

The system can only achieve high levels of utilization when the rate of data exchange between nodes handling neighboring subdomains keeps up with the computations performed on each node.  This limitation poses a challenge for strong scaling.  As a fixed-sized grid is distributed over a growing number of nodes, each processor handles fewer cells, and the ratio of communication to computation grows with the surface-to-volume ratio of subdomains.  The attainable communication to computation ratio is set in silicon during system design.  For distributed computing, this ratio is relatively low.  Communication limits result in distinct performance degradation in strong scaling plots such as those in  Fig. \ref{Strong_Scaling}, which shows a log-log plot of compute time per iteration as a function of the number of grid cells per core.  The linear region with a slope of -1 at high workload per core shows that processors remain efficiently utilized.  The distinct point where performance falls off this trend as workload per core decreases marks the end of strong scaling and the point where communication begins to dominate computation rate.


In recent decades, the gap between the computation and communication capabilities of HPC systems has been widening\cite{bib10}.  At the node level, the improvements in processing rates have outpaced the advances in memory and IO bandwidth, increasing the minimum cell count per node for balanced computations.  The problem for strong scaling is further exacerbated by cross-node interconnects whose latency and bandwidth characteristics penalize fine-grain parallelization.\cite{MVAPICH_pt2pt}  As a result, over the past two decades and for most computational fluid dynamics (CFD) applications, strong scaling has failed for workloads that use fewer than $\sim$10,000-15,000 cells per core. Further, upcoming generations of distributed HPC architectures show few signs of significant progress in strong scaling capability (See Section \ref{expl_Distr_perf} for detailed discussion).

The Wafer-Scale Engine (WSE) departs from this trend. The latest generation WSE consists of 850,000 cores manufactured on a single 12-inch wafer with over 2.6 trillion transistors. It is packaged in a system less than a meter high and a rack wide.  The WSE requires less than 1 percent of the space and power of NETL's Joule 2.0,\footnote{NETL's Joule 2.0 supercomputer is the size of a typical single family house, and consumes approximately one megawatt of power.  The WSE is packaged into a system the size of a dorm refrigerator and consumes approximately 24 kilowatts of power} yet has comparable peak performance and much higher algorithmic performance.  The ``tiles'' on the WSE minimize the fundamental components of a traditional HPC node.  Each tile has a Turing-complete, multi-gigaflop-capable processor, a local memory, and a router for communication with neighboring tiles.  Tiles form a 2D array with local communication paths between all Cartesian neighbors.  Every tile can act independently with its own program code or as a group for collective operations.  Tiles are a few hundred microns in size.  The small size allows each processor to access its own memory within a single clock cycle and a neighbor's within a single fabric hop (also single cycle).  These properties minimize latency to local and neighbor data and provide tens of petabytes per second of aggregate bandwidth.  Further, the miniaturization means that all necessary data movement is done over nanowire connections which minimizes energy consumption.\cite{Cerebras_story} 

There is no memory hierarchy on the WSE.  All memory is static random access memory local to each processor and contained within the tile.  Each processor can read 128 bits from and write 64 bits to its own memory on each cycle, and each processor can support up to eight operations per cycle depending on precision and configuration.  In addition, on each cycle, each tile can simultaneously send a 32-bit value to the interconnect and receive a 32-bit value from the interconnect.  The results in this paper show that this balanced approach to computing makes strong scaling eminently feasible for reduction-free methods and significantly improves methods with reductions.


\subsection{The Heat Equation}\label{the_diffusion_equation}
The heat equation is among the simplest field equations and follows Fourier's law as seen in Eq. \ref{diff_eq}.  Because of its simplicity, the heat equation is often used as a comparative test for field equation modeling software.
\begin{equation}\label{diff_eq}
\frac{\partial T}{\partial t} =  \alpha \nabla^2 T = \alpha \left(\frac{\partial^2 T}{\partial x^2} + \frac{\partial^2 T}{\partial y^2} + \frac{\partial^2 T}{\partial z^2}\right)
\end{equation}
The scalar field $T$ evolves in Cartesian spacetime $(x, y, z, \  \mathrm{and} \ t)$ with thermal diffusivity $\alpha$.

\subsubsection{The Explicit Formulation}\label{the_explicit_formulation}
The forward time, centered space method applied to the heat equation on a uniform Cartesian grid results in Eq. \ref{expl_form}.
\begin{equation}\label{expl_form}
\begin{gathered}
T_{C}^{n+1} = \omega \smashoperator{\sum_{D \in \mathcal N\left(C\right)}}\left(T_{D}^n\right)+ \left( 1 - 6 \omega \right) T_{C}^n \;\;\; \forall \; C \not \in \mathrm{bc}\\
T_{C}^{n+1} =T_{C}^n =\gamma \;\;\; \forall \; C \in \mathrm{bc} \\
\omega = \alpha \Delta t \Delta l^{-2}
\end{gathered}
\end{equation}

Forward time methods track a fixed spatial grid as it evolves over time. $T_{C}^n$ denotes the field $T$ in cell ${C}$ at time $n$. The next time step $T_{C}^{n+1}$ is a weighted average of a Cartesian neighborhood of the current time step. $\mathcal N\left(C\right)$ is the neighborhood operator which defines the set of cells included in the neighborhood. For first-order formulations,  $\mathcal N\left(C\right)$ includes the cells to the bottom (${B}$), top (${T}$), east (${E}$), west (${W}$), north (${N}$) and south (${S}$) of the cell under consideration. The weighting depends on diffusivity $\alpha$, time step $\Delta t$, and cell length $\Delta l$.  Cells at the boundary, $\mathrm{bc}$, of the simulation are constrained to a constant value ($\gamma$) in spacetime. The equation set is stable when the of diagonal constant ($\omega$) is less than  $\frac{1}{6}$.  All testing adopts a $\omega$ of $\frac{1}{10}$.

This simple field equation illuminates several key aspects of the hardware ecosystems used to solve it.  This equation has low arithmetic intensity.\footnote{\emph{Arithmetic intensity} is the ratio of arithmetic operations to data accesses.  The inverse ratio is \emph{data intensity}.} Processors that are separate from their main memory (e.g., with memory on physically independent RAM chips) are particularly limited in this scenario because data cannot be moved between memory and the processor fast enough to sustain peak arithmetic capabilities. Memory hierarchies such as on-chip caches can only mitigate this problem when data is reused.  For field equations, the data is only reused after the entire field is accessed. At useful field resolutions this renders caching strategies ineffective. Therefore, modern Central Processing Units (CPUs) and Graphics Processing Units (GPUs) are both limited by memory bandwidth on these problems.  In contrast, the WSE does not use a memory hierarchy.  All its memory interleaves with processing cores, allowing it to process field equations without bandwidth limitations.

\subsubsection{The Implicit Formulation}\label{the_implicit_formulation}
The backward time, centered space method applied to the heat equation on a uniform Cartesian grid results in Eq. \ref{impl_form}.

\begin{equation}\label{impl_form}
\begin{gathered}
T_{C}^{n+1} - \omega \psi \smashoperator{\sum_{D \in \mathcal N\left(C\right)}}\left(T_{D}^{n+1}\right) = \psi T_{C}^n    \;\;\; \forall \; C \not \in \mathrm{bc} \\
\psi = \left(1+6 \omega \right)^{-1}\\
T_{C}^{n+1} =T_{C}^n =\gamma \;\;\; \forall \; C \in \mathrm{bc} 
\end{gathered}
\end{equation}
\\
Eq. \ref{impl_form} represents a set of linear equations of the form $\mathbf{A}x=b$.  $\mathbf{A}$ is a sparse matrix with the characteristic sparsity pattern found in Cartesian grid decomposition\cite{hoffmann1993computational}. The diagonal is unity and non-zero off-diagonals share the same value ($-\omega$). A single multiplication of $T_{C}$ by the constant $\psi$ is the only formation operation needed as $\mathbf{A}$ is constant. The constant, $\omega$, has the same definition as in Eq. \ref{expl_form}. The Conjugate Gradient (CG) solver is applicable here as $\mathbf{A}$ is a symmetric, positive-definite matrix.\cite{daniel1967conjugate}

The CG solver is one of a class of related solvers known as Krylov Subspace solvers.  In these solvers, all-reduce operations are used to determine search step magnitudes.  In the classic CG algorithm, two reductions per iteration are needed.  They are a bottleneck at high parallelism.  Solution progress must halt while reductions are completed, which makes reduction time critical to performance for this class of solvers.  The classical CG solver iteration time was benchmarked in this study to illustrate the performance differences in reduction-dominated algorithms.

\section{WFA Programming Approach}

\begin{figure}[h]%
\centering
\includegraphics[width=1.0\textwidth]{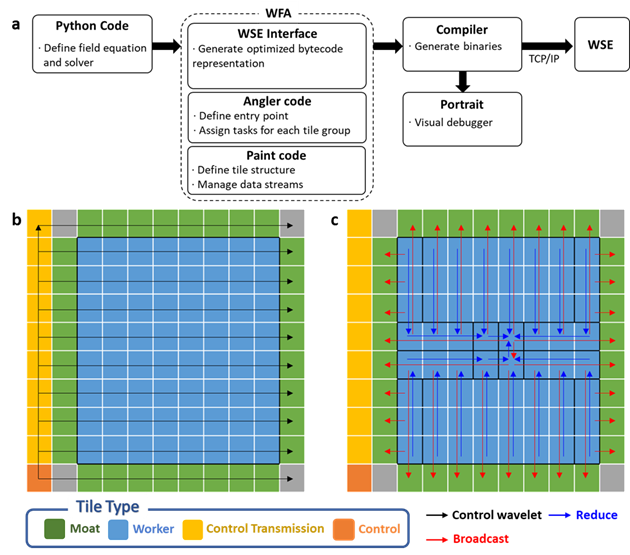}
\caption{Schematic overview of the computation strategy in the WFA (a), RPC streaming to execute vector operations in the WSE (b), and data streaming for central reductions (c).}\label{WFA_programming}
\end{figure}

The WFA offers a simple and intuitive Python user interface in a NumPy-like style.  It compiles the Python code into an executable that can run on WSE hardware and in Cerebras' visual debugger, Portrait.  Fig. \ref{WFA_programming}a outlines this workflow, and section \ref{example_code} gives examples of program code.

As shown in Fig. \ref{WFA_programming}b, the WFA organizes groups of tiles with specific functions.  To run a program, the WFA uses Remote Procedure Calls (RPCs) enacted through a bytecode sequence that is sent to an array of tiles consisting of Worker and Moat tiles. The Control Tile dynamically interprets the bytecode sequence generated at compile time, generates RPCs, and broadcasts these through the Control Transmission tiles to the Worker and Moat tiles.   Each RPC launches a coordinated set of kernels on Moats and Workers to complete a portion of a program.  Workers handle computation in the central domain.  Moats surround the Workers and handle the edge requirements of computation tasks.  These tasks could be boundary cell computations, catching data coming out of the Worker group's edge, or sending appropriate data to Workers on the edge to correctly complete a tensor operation without stalls or hangs.  The WFA has RPCs for common scalar/tensor operations and a growing library of RPCs for explicit and implicit solvers.  


This control strategy separates the main program code from the kernel code associated with RPCs (stored on Workers and Moats), which minimizes code space on the Worker and Moats. Since there is significant memory space on the Control Tile and Control Transmission tiles, large programs can fit on the WSE without need for host-WSE interaction.  Further, even larger programs can be streamed through the twelve 100Gb ethernet ports.  A single 100Gb ethernet port has enough bandwidth to support bytecode streaming for large and complex programs.  However, this strategy has yet to be developed in the WFA because the Control Tile has more than enough memory to house the bytecode sequence for a simple CFD simulation.

\subsection{Example Code}\label{example_code}

The first-generation WFA uses the domain decomposition approach shown in Fig. \ref{fig1}b. Space is discretized into a uniform Cartesian grid, and each tile handles a column of control volumes. This decomposition strategy acts as proof of concept.  Fully unstructured decompositions will be supported in future versions of the WFA.

\begin{figure}[h]%
\centering
\includegraphics[width=1.0\textwidth]{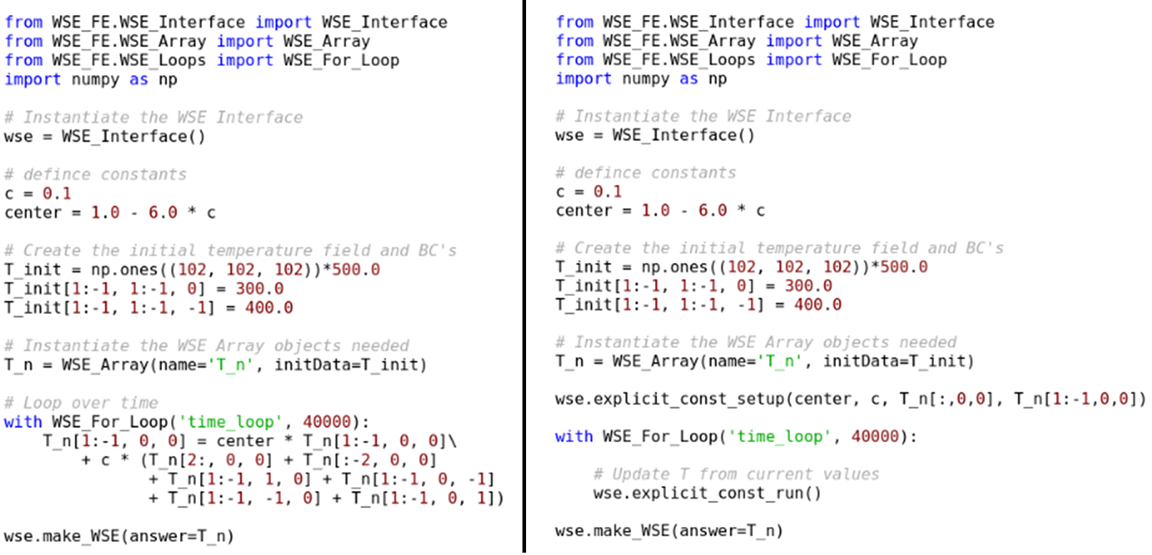}
\caption{Sample WFA code for the explicit solution of the heat equation }\label{Code_sample}
\end{figure}

Fig \ref{Code_sample} shows two example implementations of the explicit heat equation in the WFA.  The left side shows the general-purpose implementation, which specifies a \emph{for} loop with a number of vector operations that complete an explicit time step.  Notice that the definition is exactly as written in Eq. \ref{expl_form}.  

In the example on the left in Fig. \ref{Code_sample}, a temperature field is initialized in a traditional NumPy array, T\_init.  A WSE\_Array object, T\_n, is created from the initialization array.  WSE\_Array objects are very similar to NumPy arrays, but the indexing is relative to local data and processor position.  Local vector slicing is accomplished in the first axis.  The second and third axis specify relative tile position in the X and Y directions, respectively.  Values of -1, 0, and 1 in these axes specify $W$/$S$, $C$, and $E$/$N$ directions, respectively.  While the WFA does not currently understand more than these limited neighborhood specifications, there is nothing limiting the ability to specify data farther away than the immediate local neighborhood other than implementation time.  The WFA contains all the necessary functionality to write kernels for common numerical methods in Python's object-oriented programming language.

It is also possible to create complex kernels which do more than a single scalar/tensor operation.  The sample on the right does the same explicit time step in a single RPC.  In this case, it was possible to reduce overhead.\footnote{\emph{Overhead} is any computational step that is not directly an arithmetic evaluation required for the numerical method. Examples include setting pointer addresses and loop iteration counters.}  The ability to customize kernels allows users to optimize performance-critical code, implement custom communication patterns such as the central reduction shown in  Fig \ref{WFA_programming}c, and overlap communication and computation.  Custom kernels can be used together with general purpose expressions, which is valuable for making a flexible and high-performance field equation solver.

The WFA has a validation capability that can run the field equation code in NumPy and debugging capability that can be used to track data transformations on any tile (Z column in NumPy Arrays) at every operation.  This can be used to track the same data transformations in the Cerebras visual debugger, Portrait.  Finally, the entire program is compiled into WSE compatible binaries that can be shipped to an available WSE.  In this way, the ease of programmability in NumPy-like Python programming is coupled with exceptional performance that matches or exceeds the performance of past hand-optimized WSE programming efforts.\cite{bib1,bib13}

\section{Measured Performance}\label{sec4}

We compared the performance of the WFA on the WSE with that of OpenFOAM\textsuperscript{\textregistered} (hereafter referred to as OpenFOAM) on Joule 2.0.  OpenFOAM is an open-source, distributed, CFD code that is commonly used in industrial modeling and research. OpenFOAM was used as a basis for comparison because it is open-source, has single-precision support, and is often used to solve industrial problems of similar size to those possible on the WSE.  As closely as possible, the same algorithms were implemented in the WFA and in OpenFOAM. 

\begin{figure}[h]%
\includegraphics[width=0.5\textwidth]{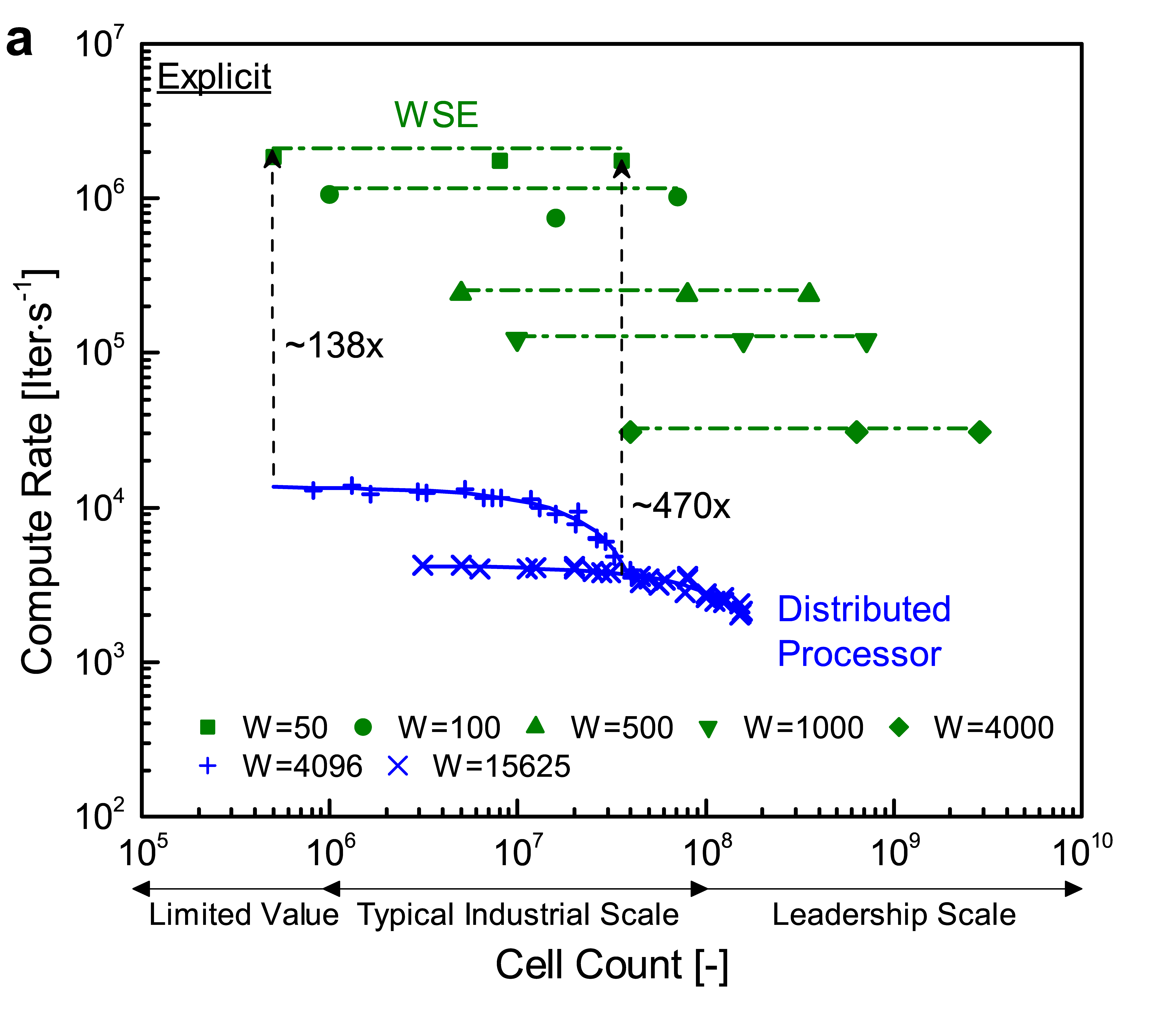}
\includegraphics[width=0.5\textwidth]{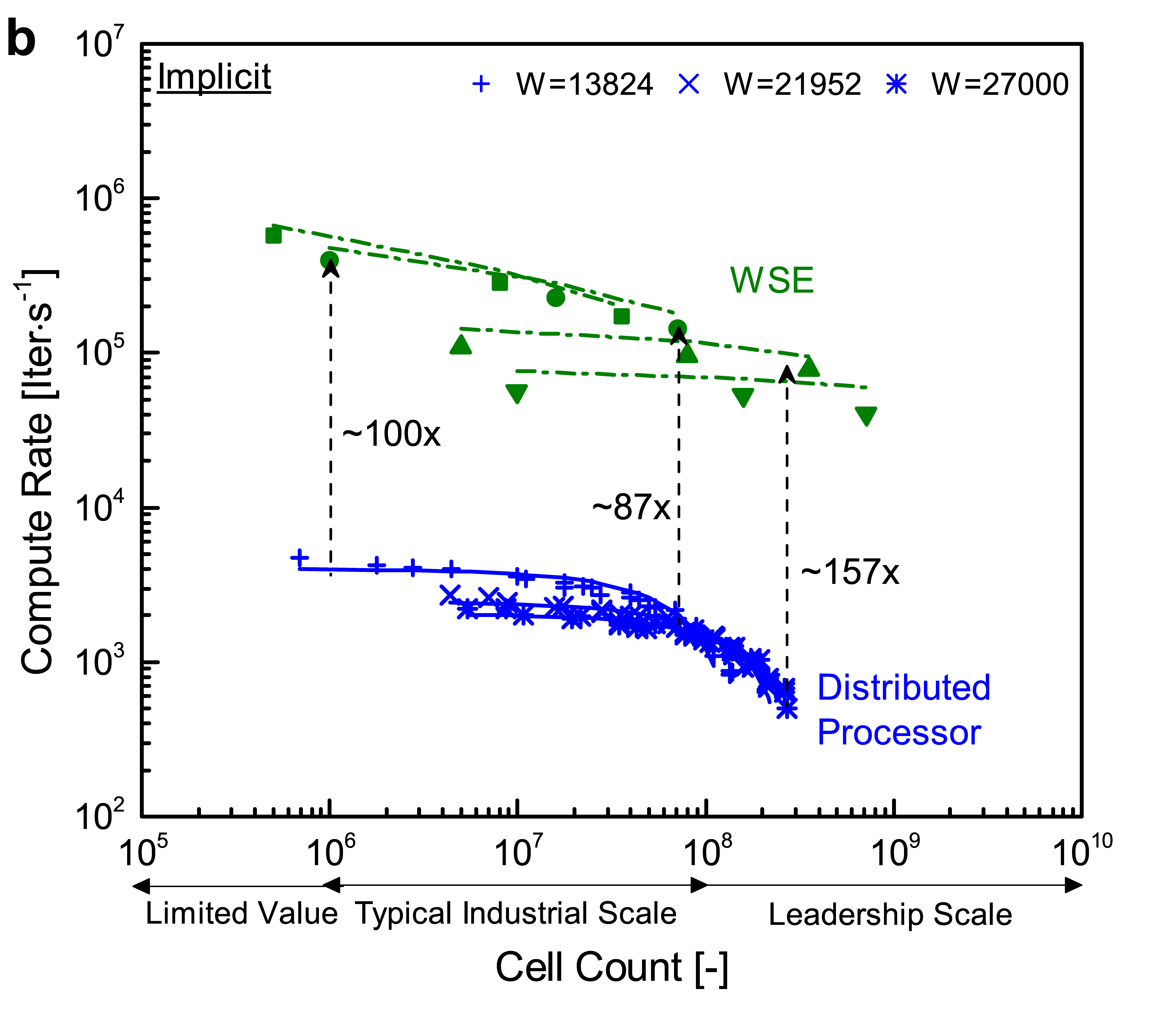}
\caption{Performance analysis of WSE (green) and distributed computing weak scaling (blue) in explicit (a) and implicit (b) formulations at various values of workload per processor (W).}\label{Perf_fig}
\end{figure}

Fig. \ref{Perf_fig} shows the performance of the WFA on the WSE against a custom-build version of OpenFOAM 8\cite{openfoam_v8,openfoam_found} on NETL's Joule 2.0 supercomputer.  Fig. \ref{Perf_fig}a shows the measured iteration speed of Eq. \ref{expl_form} for each time step.   Fig. \ref{Perf_fig}b shows the measured inner iteration speed of the classic CG algorithm.  

The code in OpenFOAM is complex, as it involves traditional halo data\footnote{\emph{Halo data} is cell data  on the periphery of the local mesh of subdomains in distributed computing that must be exchanged between neighbor processors to complete some linear algebra operations, such as a sparse, matrix-vector multiply.} exchange and all-reduce operations, expressed through the Message Passing Interface (MPI) for distributed memory parallelization.  The code in the WFA is relatively simple and can achieve parallelization on the WSE with far less code than what was required to implement an efficient distributed OpenFOAM code.  

Each result of the sparse matrix-vector product for the heat equation combines the center value and the sum of six neighboring values scaled with the diagonal and off-diagonal coefficients, respectively.  The WFA implementation uses two temporaries for this computation, one to store an element-wise product of the diagonal constant with the center value, and the other to store the sum of neighboring values.  This sum is initialized by performing an element-wise addition of locally stored top and bottom cell data.  Then, neighbor data is added to the off-diagonal sum by launching background threads that simultaneously send data to and receive data from four neighboring tiles in ${W}\rightarrow{C}\rightarrow{E}$, ${N}\rightarrow{C}\rightarrow{S}$, ${E}\rightarrow{C}\rightarrow{W}$, and  ${S}\rightarrow{C}\rightarrow{N}$ directions.  Once the sum of six neighboring values is computed, a single fused multiply-accumulate (FMAC) instruction scales it with the off-diagonal coefficient and adds it to the previously stored scaled center value.  This implementation is very lean and efficient because in Cerebras' architecture the FMAC instruction takes one cycle, and each background thread is a single machine instruction.  

In contrast, the sparse matrix-vector product in OpenFOAM cannot be briefly described at the machine instruction level.  It is implemented with coarser building blocks including: C++ libraries, MPI interfaces, and host network stack. It does local calculations on owned data, exchanges neighbor data through MPI, and then updates boundary cells with neighbor data. While these steps are similar, the implementation in code is much more complex and lengthy than what is described for the WFA above.

All-reduce operations were handled via standard calls to the MPI library in OpenFOAM.  A reduction-to-center strategy was implemented within the WFA to handle reductions as shown in Fig \ref{WFA_programming}c. 

The WSE can achieve scaling even under the worst possible communication to compute ratios.  In traditional computing, practitioners use strong scaling to find the point where communication rate fails to keep up with computation rate.  Remarkably, this never happens on the WSE, even as the ratio of communication to computation is maximized.  The sparse matrix vector multiplication (SpMV) is independent of workload per processor.  Instead, there are limits related to the overhead of setting addresses, vector lengths, and starting threads, which is expressed as the constants in the denominators of the performance models in Eqs. \ref{expl_roofline} and \ref{impl_perf_eq}. Therefore, our design is optimized for simplicity.  It uses a 1$\times$1$\times$Z domain decomposition that places every cell on four communication boundaries. The single-cycle, balanced computation and communication rate allows all vector operations to proceed at half to single cycle rates regardless of data placement. The WFA achieves perfect weak scaling in the explicit case and shows no dependency on the workload per processor as seen in Eq. \ref{expl_roofline}.  This allows strong scaling to proceed unimpeded by bandwidth and latency, which is not possible on current HPC systems.

Algorithms with reductions that can't be overlapped with computation are different (such as classic CG).  Global reductions proceed in proportion to the sum of the fabric extents (the $X+Y$ term in Eq. \ref{impl_perf_eq} as illustrated in Fig. \ref{WFA_programming}c).  Each hop costs 1 cycle.  This imposes a per-iteration latency that is independent of the workload per processor, and results in weak scaling that is not flat as processor allocation grows.

\begin{figure}[h]%
\includegraphics[width=0.5\textwidth]{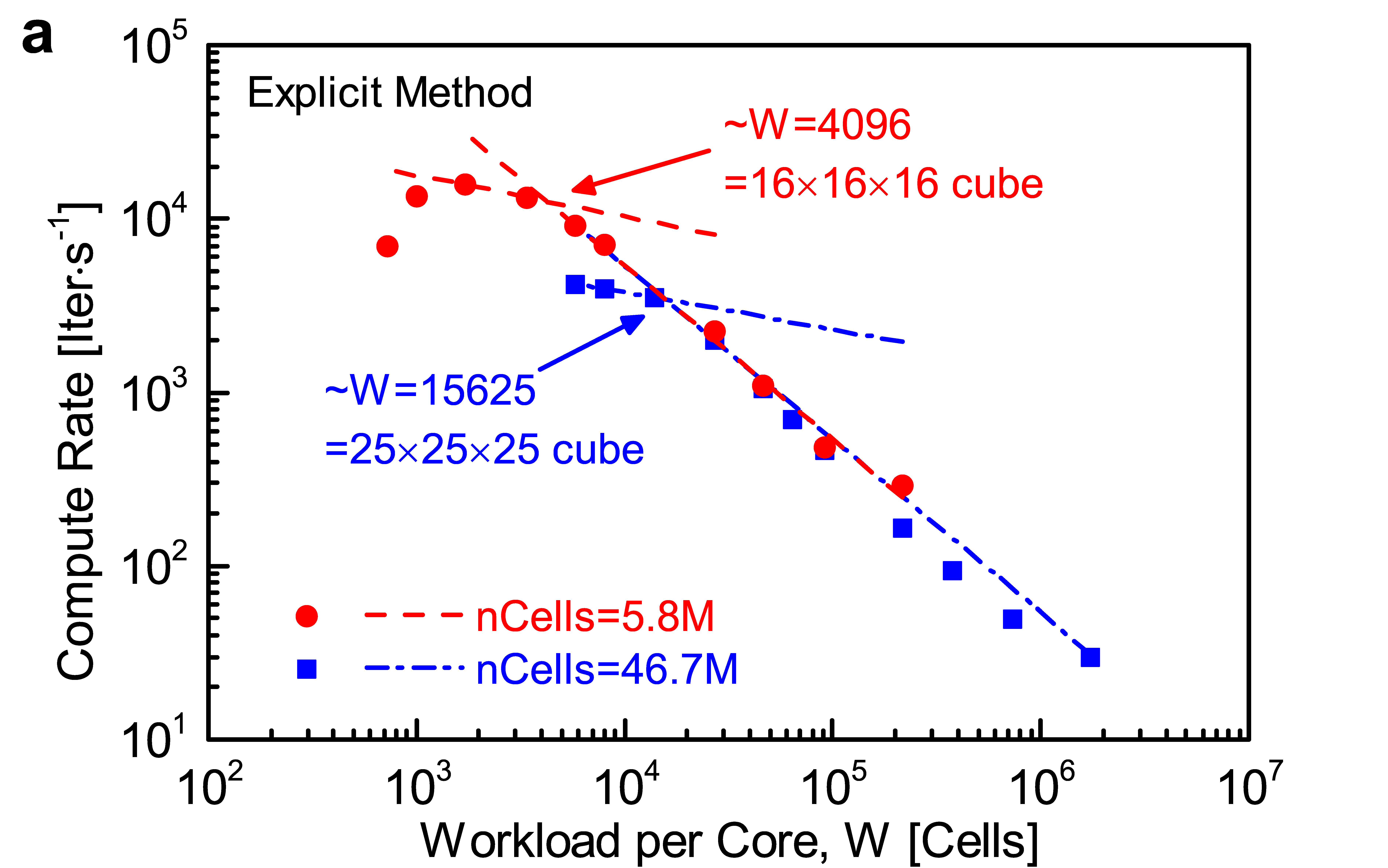}
\includegraphics[width=0.5\textwidth]{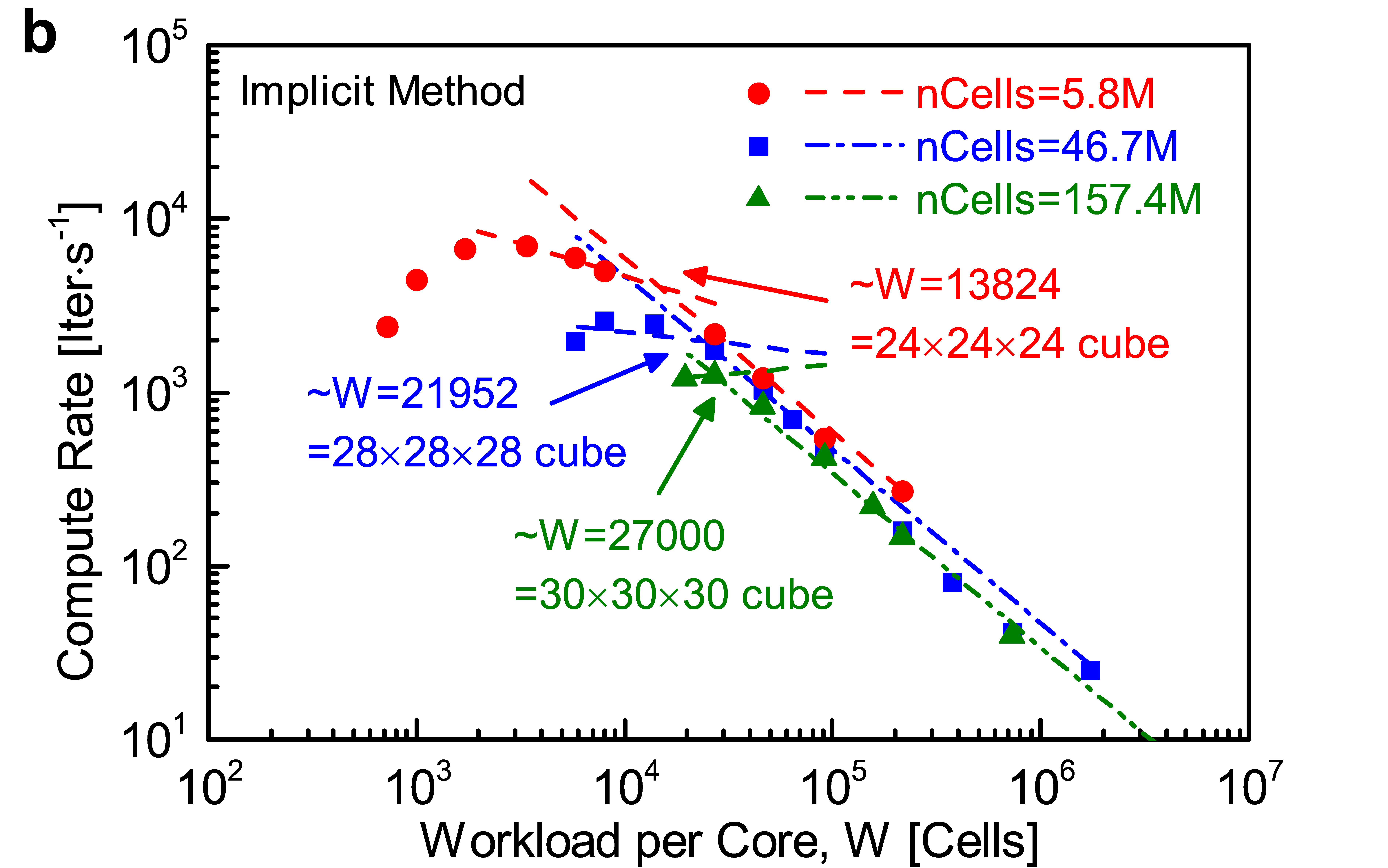}
\caption{Explicit (a) and implicit (b) strong scaling of OpenFOAM on Joule 2.0}\label{Strong_Scaling}
\end{figure}

\subsection{Explicit Formulation Performance}

Eq. \ref{expl_form} was implemented in both the WFA and OpenFOAM.  The explicit loop time was recorded for the implementations on the WSE and on Joule 2.0, then compared. 

\subsubsection{OpenFOAM on Joule 2.0}\label{expl_of_on_joule}

It is important to determine the correct workload per core, as it impacts processor utilization and achievable speed.  Workload per core is determined through strong scaling at the mesh size of interest.  The workload per processor is chosen at the point where linear strong scaling stops, as this represents the fastest iteration rate that also has high processor utilization. Strong scaling was done at both ends of the industrially relevant zone (5.8 and 46.7 million cells) as shown in Fig. \ref{Strong_Scaling}a.  This workload per processor was then used in a weak scaling study to generate competitive benchmarks (blue data in Fig. \ref{Perf_fig}a). The scaling studies used regular hexahedral cells and the ``simple'' decomposition method, which creates hexahedral subdomains.

\begin{table}[h]%
\centering
\caption{OpenFOAM Explicit Performance on Joule 2.0.}
\label{tab:explicit_of_perf}
\begin{tabular}{r|cc|cc|}
\multicolumn{1}{l|}{} & \multicolumn{2}{c|}{$W$=4,096} & \multicolumn{2}{c|}{$W$=15,625} \\
\multicolumn{1}{l|}{} & Fastest     & Slowest     & Fastest     & Slowest     \\ \hline
Iteration Speed (1/s) & 13,862      & 3,535       & 4,263       & 2,027       \\
Cell Count            & $1.31\times 10^6$ & $4.01\times 10^7$ & $5.00\times 10^6$ & $1.51\times 10^8$ \\
Core Count            & 320         & 9800        & 320         & 9680        \\
Nodes                 & 8           & 245         & 2           & 242         \\
Speed Ratio           & \multicolumn{2}{c|}{3.9}  & \multicolumn{2}{c|}{2.1} 
\end{tabular}
\end{table}

The measured performance in OpenFOAM on Joule 2.0 decays linearly with the total number of cells as seen in Eqs. \ref{of_expl_perf_small} and \ref{of_expl_perf_large} for the $W=4096$ and $W=15625$ cases, respectively.  Performance parameters are summarized in Table \ref{tab:explicit_of_perf}.  

\begin{equation}\label{of_expl_perf_small}
R_i(W=4096) = 1.36\times 10^4 - 2.55\times 10^{-4} N_c
\end{equation}
\begin{equation}\label{of_expl_perf_large}
R_i(W=15625) = 4.20\times 10^3 - 1.37\times 10^{-5} N_c
\end{equation}
\\
In the $W=4096$ case, performance dropped by 3.9 times when scaling from 320 to 9800 cores.  This is expected, given the small workload per processor.  Typical workload ranges are 10,000-15,000 cells per processor.  The $W=15625$ case scaled almost twice as well with a performance reduction of only 2.1 times from 320 cores to 9680 cores.  It is likely that explicit weak scaling could be improved further with substantial optimization efforts, but that is beyond the scope of this investigation.

\subsubsection{The WFA on the WSE}\label{wfa_on_wse}

When scaling with constant workload per processor (weak scaling), the iteration time remains constant as shown in Fig. \ref{Perf_fig}, implying that the compute rate grows linearly with the number of processors.  Using timings generated by the Cerebras hardware simulator, we developed a “roofline” model (Eq. \ref{expl_roofline}) that shows the scaling trend.   We used iterations per second as our metric.  For weak scaling, it is more important to measure the number of iterations per second (when scaling up processors at constant workload) than the rate of floating point arithmetic.  According to the calibrated model, the achievable iterations rate ($R_i$) depends only on the clock frequency ($F_c$) and the workload per processor ($W$), and not on the scale (i.e., the number of processors).  This is perfect weak scaling.

The WFA achieves high utilization on the WSE.  At these utilization levels, application performance is limited by power delivery. The single-precision WFA workload is strikingly different than the half-precision workloads typical in artificial intelligence (AI) applications, which the WSE's default power settings are tuned for.  Therefore, we tuned the power settings (including processor clock, current, and voltage) to maximize performance.   With this tuning, the actual hardware ran at speeds close to those predicted by the simulation-informed model.  We intend to integrate these power settings into a future release of the WSE platform software.  The power tuning results in an approximately 30 percent improvement in performance.


\begin{equation}\label{expl_roofline}
R_i = \frac{F_c}{6.5W + 78}
\end{equation}
\\

The WFA has the ability to scale to just 50 cells per processor while almost achieving the ideal performance in Eq. \ref{expl_roofline}.  The WSE is capable of doing 8 single precision flops of work in 6.5 cycles due to additions being done with simultaneous instruction, multiple data instructions and the FMAC.  78 cycles of overhead were involved in setting vector lengths and addresses.  It may be possible to reduce this overhead through further optimizations. By contrast, the distributed case in OpenFOAM needed over 15,000 cells to scale well.  The difference in scaling capability results in a $470\times$ improvement in iteration speed.  The WSE attains these speeds because of the balanced compute to memory and fabric access speeds, which allows it support low arithmetic intensity operations.  The stark difference in attainable speed illustrates the severe constraints imposed by the latency and bandwidth limitations in the memory and network subsystems of distributed computers during the solution of field equations.  

\subsubsection{Limits of Possible Distributed Performance}\label{expl_Distr_perf}

GPUs are the state of the art in modern HPC because the aggregate device memory bandwidth is much higher than CPUs, which can make them faster at a device level.  However, this does not necessarily translate to greater speed at the distributed system level. Fisher et. al. presented a first-principles model of distributed performance for field equation solvers.\cite{fischer2015scaling}  For explicit steps, Eq. \ref{iteration_time_model} shows Fischer's model, where the minimum time per iteration ($t_{\mathrm i}^{\mathrm{min}}$) is a function of the time to compute the inner portion of the local subdomain ($t_{\mathrm{comp}}$), the time to communicate boundary data ($t_{\mathrm{comm}}$), and the time to update the boundary cells ($t_{\mathrm b}$).

\begin{equation}\label{iteration_time_model}
t_{\mathrm i}^{\mathrm{min}} = \mathrm{max}\left(t_{\mathrm{comp}},t_{\mathrm{comm}}\right)+t_{\mathrm b}
\end{equation}
\\
Eq. \ref{iteration_time_model} models concurrent boundary data exchange with computation over the subdomains' interiors followed by computation over subdomains' boundaries.  Communication time tends to increase with the number of applied processors due to the effects of congestion, decomposition, and topology. Therefore, good weak scaling is only achieved when $t_{\mathrm{comm}} < t_{\mathrm{comp}}$ at the largest number of applied processors of interest. When this condition is met, $t_{\mathrm i}^{\mathrm{min}}$ is proportional to subdomain size and inversely proportional to memory bandwidth. In this regime, increasing memory bandwidth must be matched by a larger sub-domain size per GPU to maintain the scaling constraint ($t_{\mathrm{comm}} < t_{\mathrm{comp}}$). Over the past decade, communication times have been marginally decreasing, while memory bandwidth has been significantly increasing.  This has led to large increases in workload per GPU but little (if any) progress in time to solution.  To illustrate this point, we survey literature for similar problems, report weak scaling domain size, and calculate maximum possible iteration rate in Table \ref{gpu_weak_literature_table}. 

\begin{table}[h]%
\centering
\caption{Literature survey of weak scaling field equation solvers and estimated maximum computing rate}
\label{gpu_weak_literature_table}
\begin{tabular}{c|c|c|c|c}
\begin{tabular}[c]{@{}c@{}}\\ \\ Study\end{tabular} & \begin{tabular}[c]{@{}c@{}}Subdomain\\ Width\\ (Cells)\end{tabular} & \begin{tabular}[c]{@{}c@{}}\\W\\ (Cells)\end{tabular} & \begin{tabular}[c]{@{}c@{}}\\ \\ Processor\end{tabular} & \begin{tabular}[c]{@{}c@{}}\\$R_{\mathrm {it}}^{\mathrm {max}}$\\ (iters/sec)\end{tabular} \\ \hline
Pfister et. al. \cite{pfisterer2021gpu} & 300 & $3.28\times10^7$ & V100 & 4167 \\
Rass et. al. \cite{bib14} & 383 & $5.62\times10^7$ & P100 & 1557 \\
Rass et. al. \cite{bib14} & 512 & $1.34\times10^8$ & V100 & 838 \\
Rass et. al. \cite{bib14} & 512 & $1.34\times10^8$ & A100 & 1863 \\
Xue et. al. \cite{xue2020multi} & 256 & $1.68\times10^7$ & P100 & 5215 \\
Xue et. al. \cite{xue2020multi} & 256 & $1.68\times10^7$ & V100 & 6706 \\
Pearson et. al. \cite{pearson2020node} & 750 & $4.22\times10^8$ & V100 & 267
\end{tabular}
\end{table}

The maximum possible iteration rate, when communication time is properly hidden\footnote{\emph{Hiding communication} refers to a process where communication and computation are done in parallel and the communication time is less than the computation time.  In doing so, the communication is said to be hidden behind computation as the communication time is not time critical.}, can be determined by following the methods proposed by Rass et. al.\cite{juliaCon_2019_rass}  Their method for assessing the performance of field equation solvers is shown in Eqs. \ref{A_eff} and \ref{T_eff}. The effective memory bandwidth ($T_{\mathrm {eff}}$ in Eq. \ref{T_eff}) is related to memory footprint for unknown ($D_u$) and known ($D_k$) variables as well as the iteration time ($t_{\mathrm {it}}$).  For memory bound algorithms, this is a good metric for performance because it relates effective memory bandwidth to peak memory bandwidth.  It is also a good measure for distributed computing because $T_{\mathrm {eff}}$ will remain high when communication is properly hidden.

\begin{equation}\label{A_eff}
A_{\mathrm {eff}} = 2 D_u+D_k
\end{equation}
\begin{equation}\label{T_eff}
T_{\mathrm {eff}} = \frac{A_{\mathrm eff}}{t_{\mathrm {it}}}
\end{equation}
\\
Further, this method can estimate the maximum possible iteration rate directly from the subdomain size.  Substituting $A_{\mathrm {eff}}$ into Eq. \ref{T_eff} and solving for $t_{\mathrm {it}}$, we obtain Eq. \ref{t_it}. We now find a lower bound for $t_{\mathrm {it}}$ with aggressively optimistic assumptions. We take $D_k=0$, $D_u=1$, and assume the computation proceeds in single-precision representation while also achieving the full peak memory bandwidth  ($w_{\mathrm m}$).  This reduces iteration time to Eq. \ref{t_it_min}.  The maximum possible iteration rate is then the inverse of the iteration time (Eq. \ref{R_it_max}).

\begin{equation}\label{t_it}
t_{\mathrm {it}} = \frac{2 D_u+D_k}{T_{\mathrm {eff}}}
\end{equation}

\begin{equation}\label{t_it_min}
t_{\mathrm {it}}^{\mathrm {min}} = \frac{8W}{w_{\mathrm m}}
\end{equation}

\begin{equation}\label{R_it_max}
R_{\mathrm {it}}^{\mathrm {max}} = \left(t_{\mathrm {it}}^{\mathrm {min}}\right)^{-1}= \frac{w_{\mathrm m}}{8W}
\end{equation}
\\

The inferred maximum iteration rates in Table \ref{gpu_weak_literature_table} are all comparable to or less than those measured in OpenFOAM on Joule 2.0's CPU nodes.  The workloads per GPU are also consistent with our own experience.  There are two reasons why GPU performance is not significantly higher.  First, single device scaling on GPUs is relatively poor.  Rass et. al. report that $T_{\mathrm{eff}}$ does not peak for single-device scaling until $W$ is greater than $128^3$.  While GPU bandwidth is high, the latency is also high.  Little's law dictates that a large amount of data needs to be in flight to keep utilization high when both latency and bandwidth are high.  Sustaining significant amounts of data in flight translates to large subdomain sizes.  These single device scaling properties, limit attainable iteration rates on GPUs.  On the other hand, the WSE has L1 cache bandwidths and single cycle latency, thus the attainable iteration rates on each processor are much higher.

Second, $t_{\mathrm{comm}}$ is a not-necessarily-linear combination of all the boundary traffic communication times.  Boundary communication times are affine functions of the ratio of message size to available bandwidth and latency.  Available bandwidths tend to be low because message sizes are relatively small.  Further, latencies can be large depending on network topology, traffic, and congestion.  Traffic and congestion can have a large, negative impact on latencies and bandwidths.\cite{chunduri2019gpcnet}  These factors drive $t_{\mathrm{comm}}$ up as more processors are utilized, which can only make iteration rate drop.  We have little reason to believe that the attainable iteration rates with this specific problem will be much different than those reported in Table \ref{gpu_weak_literature_table}.\footnote{Experiments are underway on modern GPU nodes to verify workload per GPU, iteration rate, and energy consumption of the problem in Eq. \ref{expl_form} using the methods from Rass et. al.\cite{bib14,ParallelStencil_website}  These measurements are the subject of ongoing work and the findings will be released when complete.}  By way of comparison, the WSE inter-processor bandwidth matches the memory bandwidth, and inter-processor latency is single cycle.  This is necessary to sustain low intensity operations between local and neighbor data at small workloads per processor.  Further, recruiting more processors does not change anything in the communication pathways, as everything is local and no resources are shared as more cores are recruited.  These properties are foundational to the ideal weak scaling observed in the WFA in Fig. \ref{Perf_fig}a.

\subsubsection{Measured WSE Power Efficiency}

At the largest fabric sizes tested in this study ($750\times 950$), the WSE sustained an average of 24.6kW.  The computation rate per unit power was calculated to be between 32 and 35 gigaflops per watt.  This is comparable to the measured LINPACK power efficiencies reported on the Top 500 list.\cite{dongarra2003linpack,top500_website}  However, the reported values for the WFA are for real application performance, not a benchmark designed to maximally stress node level components with compute to communication ratios that are not relevant to solving field equations. It is well known that the maximum attainable flop rates in the LINPACK benchmark are close to two orders of magnitude higher than those possible in solving field equations.\cite{dongarra2016high,hpcg_website}  Unfortunately, users are not required to report power consumption for the HPCG benchmark, and Joule 2.0 does not have power monitoring equipment.  We suspect that power consumption will be close to the sum of the thermal design power limit for all processors used for explicit problems, as processor utilization is high.  Using this as a gauge, the WFA on the WSE may be more than two orders of magnitude more energy efficient than distributed computing.   Investigations are in progress to refine these estimations and will be reported when complete.  

\subsection{Implicit Formulation Performance}

Eq. \ref{impl_form} was implemented in both the WFA and OpenFOAM.  The CG inner loop time was recorded for the implementations on the WSE and on Joule 2.0.

\subsubsection{OpenFOAM on Joule 2.0}
The same strong scaling method described in \ref{expl_of_on_joule} was used to find the optimal workload per processor (Fig. \ref{Strong_Scaling}b) in the weak scaling analysis of CG (blue data in Fig. \ref{Perf_fig}b).  The optimal workload per processor for $5.8\times 10^6$ and $4.87\times 10^6$ cell simulations were 13,824 and 21,952, respectively.  In addition, there was time to run an additional strong scaling case optimized at $1.57\times 10^8$ cells with a determined workload per processor of 27,000 cells per processor.

The implicit weak scaling was also found to be near-linear as seen in Eqs. \ref{of_impl_perf_small} to \ref{of_impl_perf_large}.

\begin{equation}\label{of_impl_perf_small}
R_i(W=13824) = 3.98\times 10^3 - 2.75\times 10^{-5} N_c
\end{equation}
\begin{equation}\label{of_impl_perf_med}
R_i(W=21952) = 2.45\times 10^3 - 8.63\times 10^{-6} N_c
\end{equation}
\begin{equation}\label{of_impl_perf_large}
R_i(W=27000) = 2.05\times 10^3 - 5.66\times 10^{-6} N_c
\end{equation}
\\
The smallest implicit case has a similar workload-per-processor as the largest explicit case, but has almost half the scaling efficiency. The differences in weak scaling are also not as significant as in the explicit case.  These differences are due to reductions that occur during the two dot products in the standard CG algorithm versus the single synchronization in the explicit implementation (an MPI Waitall in the halo update).

\subsubsection{The WFA on the WSE}

The ideal performance for the CG algorithm is given in Eq. \ref{impl_perf_eq}. The performance includes overlapping the update of the solution variable, $T_c$, with the all-reduce needed to compute the norm of the residual. 

\begin{equation}\label{impl_perf_eq}
R_i = \frac{F_c}{10.5W + 2\left(X + Y\right) + 337}
\end{equation}
\\
The iteration rate was found to be not only a function of the clock frequency and workload per processor, but also the WSE fabric dimensions, $X$ and $Y$.  Performance is dominated by the size of $X$ and $Y$ when a significant portion of the WSE is allocated and $W$ is relatively small.  Unlike the explicit solver, the dependence on $X$ and $Y$ degrades weak scaling performance as wafer allocation size grows.  Each dot product in the WFA had a simulator measured performance shown in Eq. \ref{dot_perf_eq}.

\begin{equation}\label{dot_perf_eq}
t_{dot} = \frac{W + X + Y + 66}{F_c}
\end{equation}
\\
At the maximum size tested ($W=1000$, $X=750$, $Y=950$), the dot product can be done in 3.25$\mu$s.  This is substantially lower than most measured distributed computing reductions.  The MVAPICH group reported all-reduce latencies for 1024 CPU nodes between 15 and 35$\mu$s with and without SHArP \cite{MVAPICH_collectives}.  SHArP is an in-network computing technology available on InfiniBand networks that is designed to reduce  collective latency.\cite{graham2020scalable, infiniband_ta, infiniband_nvidia} Typical GPU all-reduce latencies are over 100$\mu$s without SHArP, and it has been reported that SHArP can double reduction bandwidth \cite{gtc_song}.  All-reduce latency is critical to total system performance, especially in algorithms where computation and all-reduce cannot be overlapped.  The disparity between the WSE and distributed all-reduce contributes to the performance advantage.

The measured iterative performance on the WSE was approximately 7.7 times lower than the explicit solver at full fabric allocation with small $W$, even though the implicit solver has only twice as many flops (8 vs 15).  When the WSE is capable of doing all the work outside of the reductions in close to a microsecond, even a small amount of time ($\approx$6.5 microseconds) spent on two dot products has a large impact on achievable performance when communication cannot be overlapped with compute.  It may worth exploring reduction-free and/or reduction-limited implicit methods as they will likely yield higher performance gains.\cite{bib14,cools2017communication,ghysels2014hiding}  However, the iterative convergence rate must be assessed against the iteration time to decide which yields better overall performance.

\section{Impact on Science and Technology}

Should this approach be adopted and democratized, the impact on science and technology would be significant.  The technology would help facilitate design and engineering, optimize equipment operations, and boost the rate of scientific progress.  Field equation models are widely used in industry to design and engineer products before production.  The most common use of field equation models is in system simulations for design optimization and uncertainty quantification.  Currently, the limits in time-to-solution often force practitioners to examine fewer design points or make model simplifications to complete design space explorations in a reasonable time frame.  Practitioners could use this technology to significantly speed preproduction design activities and/or ensure a more thorough exploration of the design space, which could result in more optimized designs, reduced risks, and greater certainty.  For postproduction operations, the speed gains are significant enough to allow high-fidelity models to run at or faster than real-time.  This allows detailed scientific models to be used as digital twins for equipment operations and could enable transformative technology in the areas of cyber-physical security, equipment degradation monitoring/prediction, real time command and control, flexible operations, and upset recovery.  In regard to research productivity, much of humanities modern understanding of science and the universe comes from developing, testing, and refining complex models.  Current simulations of physical systems are slow and tedious, which has a significant impact on the rate at which humankind can generate novel ideas.  A significant loosening of this bottleneck will proportionately increase the rate at which researchers can generate and apply new scientific knowledge.

\section{Conclusions}

With a Python front end that can digest high-level tensor expressions directly into compiled code to run on the WSE, the WFA provides WSE users with a simple and intuitive programming interface.  The simplicity of the approach should not be ignored.  In less than 18 months, NETL was able to develop a new programming methodology with a team of three people on never-before-seen hardware with a unique instruction set.  In contrast, efficient distributed computing efforts often take years to decades of work with very large groups of developers.  Optimization of distributed codes is a substantial undertaking and has a high cost in time and labor.  This is due in large part to the complexities of overlapping computation with communication and  managing complex memory hierarchies and complex network topologies.   Codes that can do this well and maintain generality tend to be very big and difficult to use or understand.  The WFA is much simpler.  The simplicity and low development time are the result of the straightforward WSE architecture, especially the absence of a memory hierarchy, the ideal ratio of computing rate to memory bandwidth, the inherent support for tensor instructions, and use of a single device that can achieve useful scales.  These properties give the WSE an inherent parallelism that significantly reduces development time and complexity.  While it is true that these properties end at the wafer's edge, and that there are limits to the available memory, the computing concept should not be ignored, but rather developed further.  The value of the WFA on the current generation WSE is in solving medium to large scale industrial problems.  The attainable grid size on the WSE is comparable to typical large industrial scales run on mid-size clusters, yet the solution speed can be more than two orders of magnitude faster.  Because of this, NETL is investing in developing the WFA further to support finite volume CFD methods.

The WFA on the WSE solved the explicitly formulated heat equation as much as 470 times faster than OpenFOAM on Joule 2.0.  The performance leap illustrates how much performance is lost due to latency and bandwidth limitations in the memory and network subsystems of distributed computing architectures.  The balanced computing rate to memory access ratio on the WSE architecture supports low arithmetic intensity operations well, which extends strong scaling to much lower workloads per processor.  In combination with the substantial array of tiles, the current generation WSE is valuable for solving industrially relevant problems.  The WFA on the WSE was also able to perfectly weak scale out to as many as $2.85\times 10^9$ cells, a capacity usually only seen on large supercomputers.

The CG implementation in the WFA outperformed the CG implementation in OpenFOAM by a factor of at least 87 at large industrial scales.  The iterative performance of the implicit solver was approximately 7.7 times lower than the explicit solver on the WSE at full fabric allocation with small $W$, even though the implicit solver has only twice the work to do.  This illustrates just how powerful the WSE is in low arithmetic intensity operations.  The reductions times on the WSE are over an order of magnitude faster than what is possible on distributed computing, yet the time to compute everything outside the dot product is over 400 times faster.  Thus, reductions have a proportionately larger impact on WSE performance as compared to distributed computing. Nevertheless, the WFA on the WSE outperformed OpenFOAM on Joule 2.0 by nearly two orders of magnitude with the classic CG algorithm.  It is likely that pipelined versions of Krylov subspace solvers will see larger gains, and new reduction-free implicit methods will be close to the performance gains in the explicit formulation.

\section{Obtaining the Code}

The WFA is open source and available by request on NETL's private gitlab server.  Instructions and documentation for obtaining the code can be found \href{https://dirk-netl.github.io/WSE_FE/}{here}.\cite{WFA_documentation_website}  The source code for the OpenFOAM benchmarks can be found \href{https://github.com/dirk-netl/OpenFoamBenchmarks}{here}.\cite{OF_benchmark_code}.  Instructions for compiling and running the benchmarks are given in the README file.

\section{Disclaimer}

This project was funded by the Department of Energy, National Energy Technology Laboratory an agency of the United States Government, through an appointment administered by the Oak Ridge Institute for Science and Education. Neither the United States Government nor any agency thereof, nor any of its employees, nor the support contractor, nor any of their employees, makes any warranty, expressor implied, or assumes any legal liability or responsibility for the accuracy, completeness, or usefulness of any information, apparatus, product, or process disclosed, or represents that its use would not infringe privately owned rights.  Reference herein to any specific commercial product, process, or service by trade name, trademark, manufacturer, or otherwise does not necessarily constitute or imply its endorsement, recommendation, or favoring by the United States Government or any agency thereof. The views and opinions of authors expressed herein do not necessarily state or reflect those of the United States Government or any agency thereof.


\bibliography{main}


\end{document}